# Room-temperature perpendicular magnetization switching through giant spin-orbit torque from sputtered $Bi_xSe_{(1-x)}$ topological insulator material

Mahendra DC[1], Mahdi Jamali[2], Jun-Yang Chen[2], Danielle Reifsnyder Hickey[3], Delin Zhang[2], Zhengyang Zhao[2], Hongshi Li[3], P. Quarterman[2], Yang Lv[2], Mo Li[2], K. Andre Mkhoyan[3] and Jian-Ping Wang[2,1,3],*

[1]*School of Physics and Astronomy, University of Minnesota, MN 55455*
[2]*Department of Electrical and Computer Engineering, University of Minnesota, MN 55455*
[3]*Department of Chemical Engineering and Material Science, University of Minnesota, MN 55455*

The spin-orbit torque (SOT) arising from materials with large spin-orbit coupling promises a path for ultra-low power and fast magnetic-based storage and computational devices. We investigated the SOT from magnetron-sputtered $Bi_xSe_{(1-x)}$ thin films in $Bi_xSe_{(1-x)}$/CoFeB heterostructures by using a dc planar Hall method. Remarkably, the spin Hall angle (SHA) was found to be as large as 18.83, which is the largest ever reported at room temperature (RT). Moreover, switching of a perpendicular CoFeB multilayer using SOT from the $Bi_xSe_{(1-x)}$ has been observed with the lowest-ever switching current density reported in a bilayer system: $2.3 \times 10^5$ A/cm$^2$ at RT. The giant SHA, smooth surface, ease of growth of the films on silicon substrate, successful growth and switching of a perpendicular CoFeB multilayer on $Bi_xSe_{(1-x)}$ film opens a path for use of $Bi_xSe_{(1-x)}$ topological insulator (TI) material as a spin-current generator in SOT-based memory and logic devices.

**Summary Sentence: We demonstrated growth of smooth $Bi_xSe_{(1-x)}$ films on a large silicon wafer with the largest spin Hall angle ever reported using a semiconductor industry compatible sputtering process; furthermore, we developed and switched a perpendicular CoFeB multilayer on $Bi_xSe_{(1-x)}$ films for the first time at RT by $Bi_xSe_{(1-x)}$, a TI material, which enables a path for reliable and efficient beyond-CMOS devices.**

*Corresponding author: jpwang@umn.edu

Currently, SOT in spin Hall material (SHM)/ferromagnet (FM) heterostructures is of great interest due to its efficient switching of magnetization in proposed spin-based memory and logic devices (*1–3*). SOT has been calculated theoretically (*4*, *5*) and observed experimentally in heavy metals (HMs)(*6–10*) and TIs (*11–16*). The in-plane charge current injected into an SHM/FM/oxide heterostructure is scattered in the vertical direction, either up or down, depending upon the spin orientation of electrons due to the spin Hall effect (SHE) (*6*, *7*, *10*, *17*) or Rashba effect (*9*, *18*). Thus, accumulated spin-polarized current in the interface between the SHM and FM transfers spin angular momentum to the FM (*6*, *7*, *19*), which can rotate the magnetization of the FM and is known as SOT. Conventionally, the spin-polarized current is generated by using a ferromagnetic polarizer that transfers its spin angular momentum to a FM layer. Spin-transfer torque (STT) based devices suffer from large power consumption and reliability issues due to the low efficiency of the ferromagnetic polarizer (*6*, *20*). SOT-based memory and logic devices are superior compared to the STT-based devices because they do not require a separate polarizer for the generation of spin-polarized current and could potentially have a much more efficient spin-current source (*1*, *6*, *10*). This results in a lower writing current density and, thus, much better device reliability.

The most commonly studied spin-current generators in SHM/FM heterostructures are HMs, such as Ta (*6*, *8*, *21*), W (*10*, *17*), Pt (*7*, *9*), and TIs such as $Bi_2Se_3$(*11*, *12*, *14*), and $(Bi_{0.5}Sb_{0.5})_2Te_3$ (*13*). It has been reported by several groups that a large current density on the order of $10^6$-$10^8$ A/cm$^2$ is required to switch the magnetization using SOT from HMs due to their small SHA (*6*, *7*, *9*). In the case of a TI as the spin-current generator, switching of the magnetic doped TI $(Cr_{0.08}Bi_{0.54}Sb_{0.38})_2Te_3$ layer at 1.9 K has been observed with a current density $8.9 \times 10^4$ A/cm$^2$ (*13*). It is also expected that TIs will be able to demonstrate a low switching current density for the switching of magnetization at RT as a result of their large spin-orbit coupling (SOC) (*11–14*). However, the switching of magnetization has not yet been observed on any TI/FM bilayer system at RT. Furthermore, the practical application of single crystalline TI grown by molecular beam epitaxy (MBE) is limited due to its strict demand on the single crystal substrate. It is also constrained due to the presence of valleys or voids with different size and height as a result of triangular spirals (*22*).

In this paper, we report $Bi_xSe_{(1-x)}$ films with giant spin Hall angles at RT grown on silicon substrate by magnetron sputtering, which is a semiconductor industry compatible process. The dc planar Hall method is used for the characterization of SOT in a $Bi_xSe_{(1-x)}$/CoFeB heterostructure with in-plane magnetic easy-axis (*23–25*). At RT, the SHA (the ratio of spin-polarized current density to charge current density) of the sputtered $Bi_xSe_{(1-x)}$ film is up to two orders of magnitude larger than that of HMs and approximately one order of magnitude greater than the crystalline TI, $Bi_2Se_3$. In particular, we developed a perpendicular CoFeB multilayer on $Bi_xSe_{(1-x)}$ films and demonstrated switching of the magnetization using SOT arising from the $Bi_xSe_{(1-x)}$ with the lowest switching current density in bilayers at RT. Moreover, the spin Hall conductivity (SHC), which is the product of SHA and electrical conductivity of the SHM, is determined to be comparable to or larger than the previously reported values of other spin-current generators. Furthermore, the sputtered $Bi_xSe_{(1-x)}$ layer on silicon substrate shows an excellent surface smoothness on a wafer scale for a better practical application in future.

In order to characterize the SOT arising from the $Bi_xSe_{(1-x)}$ films, thin films with the multilayer structure Si/SiO$_2$/MgO (2 nm)/$Bi_xSe_{(1-x)}$ ($t_{BiSe}$ nm)/CoFeB (5 nm)/MgO (2 nm)/Ta (5 nm) were prepared, with $t_{BiSe}$= 4, 8,16, and 40 nm as shown in the schematic drawing in figure 2A. Unless otherwise stated, we will use the labeling BS4, BS8, BS16, BS40 for the samples with $t_{BiSe}$= 4, 8, 16, 40 nm, respectively. Figure 1A,B shows the high-angle annular dark-field scanning transmission electron microscopy (HAADF-STEM) images of samples BS4 and BS8, respectively. The HAADF-STEM images show that $Bi_xSe_{(1-x)}$ has a polycrystalline structure and that the atomic layers of Bi and Se are continuous in both the BS4 and BS8 samples. Additionally, the average grain orientation in sample BS4 is 2° with a standard deviation of 9° (from vertical c-orientation), which is almost identical to the average grain orientation in the BS8 sample (2° with a standard deviation of 8°). The energy dispersive X-ray spectroscopy (EDS) line scan shows that at the top of the $Bi_xSe_{(1-x)}$ films there is stoichiometric $Bi_2Se_3$; however, there is a gradient of the Bi concentration from top to the bottom of the films (figure S2). The average ratio x in $Bi_xSe_{(1-x)}$ film is 0.47 with ± 3% uncertainty determined by the Rutherford back scattering (RBS). Figure 1C shows the atomic force microscopy (AFM) images of $Si/SiO_2/MgO$ (2 nm)/$Bi_xSe_{(1-x)}$ 4 nm film. The root mean square (RMS) value of the surface roughness of the 4 nm $Bi_xSe_{(1-x)}$ film is 0.51nm. Furthermore, we probed surface roughness propagation in full stack as shown in figure 1D. The RMS value of surface roughness 0.38 nm in BS4 sample confirms the smoothness of the full stack for future device fabrication on a wafer level.

The multilayer films are patterned into Hall cross bars 5-30 μm wide and 70 μm long, and dc planar Hall measurement is performed on the Hall cross bar with dimensions 10 μm × 70 μm, as shown in figure 2B. The bipolar input current of magnitude up to 8.5 mA is injected along the *x*-direction, and the angle-dependent Hall resistance ($R_H(I,\alpha)$) is measured under the application of a constant 0.5 T in-plane magnetic field while rotating the sample in the *xy* plane from -7 to 365°. The in-plane SOT exerted by the accumulated spin current on the interface of the $Bi_xSe_{(1-x)}$/CoFeB (figure 2A) is obtained by using Slonczewski's equation (*19*), $\tau_\parallel = \frac{\hbar J_s}{2eM_s t}(\hat{m} \times (\hat{\sigma} \times \hat{m}))$, where $\hbar$ is the reduced Planck's constant, $J_s$ is the spin-polarized current density, $e$ is an electronic charge, $M_s$ is the saturation magnetization, *t* is the thickness of the ferromagnetic layer, $\hat{m}$ is the magnetization unit vector, and $\hat{\sigma}$ is the spin-polarized current accumulation unit vector. The associated out-of-plane magnetic field with the in-plane torque ($\tau_\parallel$) is given by (*6*) $H_{OOP} = \frac{\hbar J_s}{2eM_s t}(\hat{\sigma} \times \hat{m})$. In addition to the in-plane torque, there is also an out-of-plane component of SOT due to the spin accumulation at the interface, which is given by $\tau_\perp = \alpha_T(\hat{m} \times \hat{\sigma})$, where $\alpha_T$ is a coefficient that determines the efficiency of current-induced effects. The associated in-plane magnetic field with the out-of-plane torque ($\tau_\perp$) is given by $H_T = \alpha_T \hat{\sigma}$. Figure 2C shows $R_H(I,\alpha)$ for the sample BS4 at ± 8.5 mA. Moreover, $R_H(I,\alpha)$ shows two-fold symmetry with extrema at approximately 45° in increments of 90°. In principle, the $R_H(I,\alpha)$ consists of the planar Hall resistance ( $R_{PHE}$) and the anomalous Hall resistance ($R_{AHE}$) due to the planar and anomalous Hall effects, respectively. The $R_{PHE}$ is due to the combined effects of the external field, current-induced effective fields, and the anisotropy field acting on the magnetization ($R_{PHE} \propto \sin(2\beta)$, where $\beta$ is the angle between the magnetization and current flow direction). The $R_{AHE}$ is due to the pulling of the magnetization in the out-of-plane direction by a current-induced effective field ($R_{AHE} \propto M_z$, where $M_z$ is the z-component of magnetization).

The current-induced effective fields $H_T$ and $H_{OOP}$ can be extracted by characterizing the $R_H(I,\alpha)$ for positive and negative input currents. The difference of the Hall resistance is given by (*23*),

$$R_{DH}(I,\alpha) = R_{PHE}(I,\alpha) - R_{PHE}(-I,\alpha) + 2\frac{dR_{AHE}}{dH} H_{OOP} \cos\alpha + C \qquad (1),$$

where $C$ is the resistance offset that accounts for the Hall bar imbalance, $\frac{dR_{AHE}}{dH}$ is the gradient of the change in the anomalous Hall signal versus the externally applied out-of-plane magnetic field, and $\alpha$ is the angle between field and current flow direction. Figure 2D shows $R_{DH}(I,\alpha)$ versus an externally applied field angle for the sample BS4 at different input currents. The $R_{DH}(I,\alpha)$ increases with increase in input current and has a maximum located at about $180°$. The $\frac{dR_{AHE}}{dH}$ is obtained by sweeping the out-of-plane field at a small constant input current. After considering current shunting and short circuit effects (*26*), the $\frac{dR_{AHE}}{dH}$ is determined to be - 6.75, 7.33, -7.37, -7.55, -6.51 for BS4-BS40 samples, respectively. After determining the $\frac{dR_{AHE}}{dH}$ values, $H_{OOP}$ can be determined by curve fitting of the $R_{DH}(I,\alpha)$ formula over the experimental data. The $H_{OOP}$ versus the current density is presented in figure 2E. The $\frac{H_{OOP}}{J_{SHM}}$ determined by the linear fit is as large as (99.94 ± 0.17) Oe ($10^{-6}$)/ (A/cm²) for the BS4 sample, where $J_{SHM}$ is the current density in the Bi$_x$Se$_{(1-x)}$ layer (the uncertainty is the standard error from the linear fit). The SHA ($\theta_{SH} = \frac{2eM_s t}{\hbar} \frac{H_{OOP}}{J_{SHM}}$) of the BS4 sample is determined to be 18.83 ± 0.03. Sample BS4 has the largest SHA value at RT that has been reported to date, which means that it is the most efficient spin-current source at RT. The $\frac{H_{OOP}}{J_{SHM}}$ for BS6-BS40 samples is (23.17 ± 0.48), (14.19 ± 0.24), (9.27 ± 0.03), and (2.14 ± 0.11) Oe (10$^{-6}$)/(A/cm$^2$), respectively. The SHAs for BS6-BS40 samples are determined to be 4.36 ± 0.09, 2.67 ± 0.04, 1.74 ± 0.05, 0.4 ± 0.02, respectively. The SHC is defined as $\frac{\hbar}{2e}\theta_{SH}\sigma$, where $\sigma$ is electrical conductivity of the SHM. The SHC is determined to be as large as $1.47 \pm 0.02 \times 10^5 \frac{\hbar}{2e}$ $\Omega^{-1}$m$^{-1}$ for the BS4 sample. The Bi$_x$Se$_{(1-x)}$ films have both $\sigma$ and SHC values comparable to previous reports on TI (*11–14*). A summary of $\sigma$, SHA, SHC for our samples and the best previously reported SHMs is presented in Table 1.

In our measurement technique, after we use a large in-plane magnetic field to make the CoFeB film single domain, we did not observe any significant $H_T$ and Oersted field which can be clearly seen in the $R_{DH}(I,\alpha)$ signal that has perfect $\cos\alpha$ nature (figure 2D). This indicates that the large SOT in Bi$_x$Se$_{(1-x)}$ films is due to the SHE. The origin of SHE in our Bi$_x$Se$_{(1-x)}$ films is investigated by studying variations of $\sigma_{SH}$ with $\sigma$. Analogous to the scaling law of AHE in the FM films (*27, 28*), the SHC can be written as $\sigma_{SH} = \sigma_{int} - \delta\sigma^2$, where $\sigma_{int}$ is the intrinsic SHC and $\delta$ is a constant which determines the contribution of the extrinsic SHE. The $\sigma_{SH}$ versus the square of electrical conductivity is presented in figure 2E. By fitting the above equation over the experimental data, $\sigma_{int}$ and $\delta$ are determined to be (-1.49 ± 0.11) × 10$^5$ $\frac{\hbar}{2e}$ $\Omega^{-1}$m$^{-1}$ and (-4.07 ± 0.08) $\frac{\hbar}{2e}$ $\Omega^{-1}$m$^{-1}$, respectively. The dominant $\sigma_{int}$ confirms that the SHE in the Bi$_x$Se$_{(1-x)}$ films is mainly due to the intrinsic SHE.

The SOT arising from $Bi_xSe_{(1-x)}$ can be directly observed by switching a FM in close proximity to the spin channel(*6–9*). We prepared a Si/SiO$_2$/MgO (2 nm)/$Bi_xSe_{(1-x)}$ (4 nm)/Ta (0.5 nm)/CoFeB (0.6 nm)/Gd (1.2 nm/CoFeB (1.1 nm)/MgO (2 nm)/Ta (2 nm) sample and a control sample Si/SiO$_2$/MgO (2 nm)/Ta (5 nm)/CoFeB (0.6 nm)/Gd (1.2 nm)/CoFeB (1.1 nm)/MgO (2 nm)/Ta (2 nm) for the switching experiment. The films were patterned into Hall cross bars using optical lithography. Figure 3B shows the $R_{AHE}$ loop of the $Bi_xSe_{(1-x)}$ switching sample obtained by sweeping the out-of-plane field at a constant input current of 50 µA. The non-zero $R_{AHE}$ at zero magnetic field confirms the easy axis of the magnetization along the out-of-plane direction. The perpendicular magnetic anisotropy (PMA) originated from the exchange interaction between the CoFeB and Gd layers. The $R_{AHE}$ loops of the $Bi_xSe_{(1-x)}$ switching sample as a result of dc current sweep under the application of a constant in-plane bias field along the current channel are shown in figure 3C,D. In the presence of a + 80 Oe bias field (figure 3C), the dc current sweep from positive to negative favors the magnetization in the downward direction ($M_z<0$), and the magnetization switching occurs at -7 mA (~2.3 × 10$^5$ A/cm$^2$). The subsequent reverse sweep from negative to positive current at identical magnitude and direction of the field favors the magnetization in the upward direction ($M_z>0$), and the magnetization switching occurs at 7 mA. Upon changing the polarity of the external field, the chirality of the $R_{AHE}$ loop changes, which is consistent with the results of previous reports (*6–8*). The steps in switching confirm that the switching is mainly due to the domain wall nucleation and domain wall motion. The orientation of the $R_{AHE}$ loop in the Ta switching sample is opposite to the $Bi_xSe_{(1-x)}$ switching sample, which suggests that the two samples do have opposite SHA polarity (figures 3C,D, and S5C,S5D). We also calculate the switching efficiency (*29*),

$\eta = \dfrac{H_k - H_x}{J_{sw}}$, where $H_k$ is the anisotropy field, $H_x$ is the external bias field, and $J_{sw}$ is the switching current density. The switching efficiency of the $Bi_xSe_{(1-x)}$ is determined to be 2.57 kOe per 10$^5$ A/cm$^2$. The Ta has switching efficiency of 4.7 kOe per 10$^7$ A/cm$^2$, which is comparable to the previous reports (*29*). The two orders of magnitude larger switching efficiency and opposite $R_{AHE}$ loop orientation in the $Bi_xSe_{(1-}$ $_{x)}$ switching sample compared to those of the Ta switching sample confirm that the switching is due to SOT from the 4 nm thick $Bi_xSe_{(1-x)}$ layer rather than the 0.5 nm thick Ta insertion layer in the $Bi_xSe_{(1-x)}$ switching sample.

One of the key factors for widespread realization of SOT-based spintronic devices is the development of a spin Hall channel with a high spin current generation efficiency. This leads to a low switching current density of the adjacent magnet and paves the way for ease of integration of spintronic devices with semiconductors. Moreover, the development of a smooth SHM interface at low thermal budget must be achievable. We demonstrated the growth of smooth $Bi_xSe_{(1-x)}$ films on a large silicon wafer with the largest SHA ever reported using a semiconductor industry compatible sputtering process. These films possess comparable or better SHC compared to other reported spin-current generators at RT. Furthermore, we developed and switched a perpendicular CoFeB multilayer on $Bi_xSe_{(1-x)}$ films for the first time at RT by a TI material, which enables a path for reliable and efficient beyond-CMOS devices. It should be pointed out that the $Bi_xSe_{(1-x)}$ in this study is still not optimized, so there is a wide window to improve the characteristics of $Bi_xSe_{(1-x)}$ as a spin Hall channel.

**ACKNOWLEDGMENTS**

We would like to thank Prof. Paul Crowell for proofreading the manuscript and Prof. M. Kawaguchi for helpful discussions on data analysis. **Funding:** This work was supported by C-SPIN, one of six STARnet program research centers. This work utilized (1) the College of Science and Engineering (CSE) Characterization Facility, University of Minnesota (UM), supported in part by the NSF through the UMN MRSEC program (No. DMR-1420013); and (2) the CSE Minnesota Nano Center, UM, supported in part by NSF through the NNIN program.

**Fig. 1. STEM and surface roughness characterization. (A)** and **(B)** Composites of simultaneously acquired BF- and HAADF-STEM images of samples BS4 and BS8, respectively. The selected region of the HAADF-STEM image indicated by a black line is magnified at right to show the $Bi_xSe_{(1-x)}$ atomic detail. The AFM images of **(C)** 4 nm $Bi_xSe_{(1-x)}$ film and **(D)** BS4 sample.

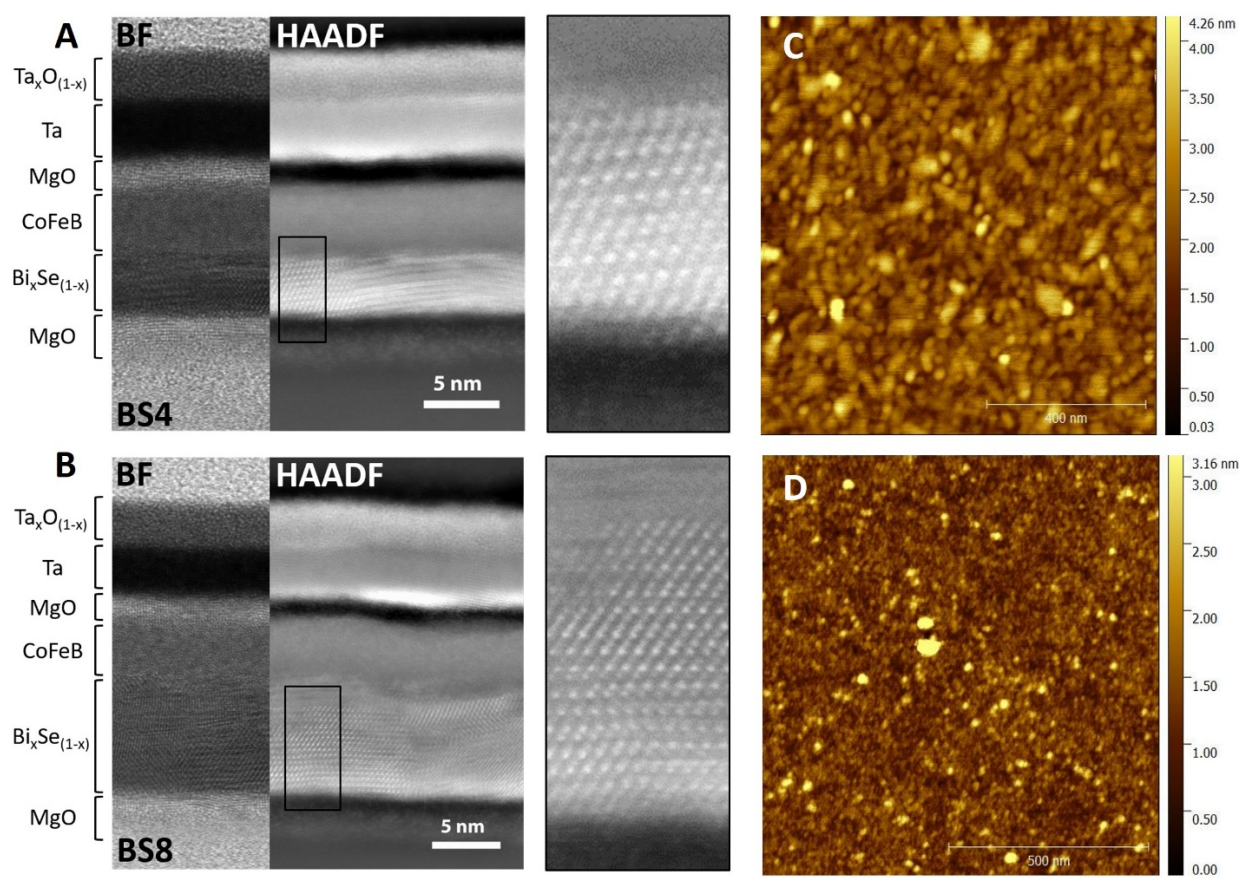

**Fig. 2. Schematic diagram, experimental set up, angle-dependent Hall resistance measurements, and characterization of SOT.** (**A**) A three-dimensional schematic diagram demonstrating the SOT in a $Bi_xSe_{(1-x)}$/CoFeB heterostructure. $H_{ext}$ and *M* represent an in-plane externally applied magnetic field and the in-plane magnetization, respectively. $H_T$ and $H_{OOP}$ are the current-induced transverse and out-of-plane magnetic fields, respectively. (**B**) An optical micrograph of the fabricated Hall cross bar with schematic drawings of the Hall measurement set up. (**C**) The $R_H(I,\alpha)$ of sample BS4 at ± 8.5 mA input current on the left axis and $R_{DH}(I,\alpha)$ on the right axis at RT under a constant 0.5 T in-plane magnetic field. (**D**) The $R_{DH}(I,\alpha)$ at different input currents for the BS4 sample. (**E**) The $H_{OOP}$ variation with the current density. (**F**) The $\sigma_{SHC}$ change with respect to the linear conductivity squared.

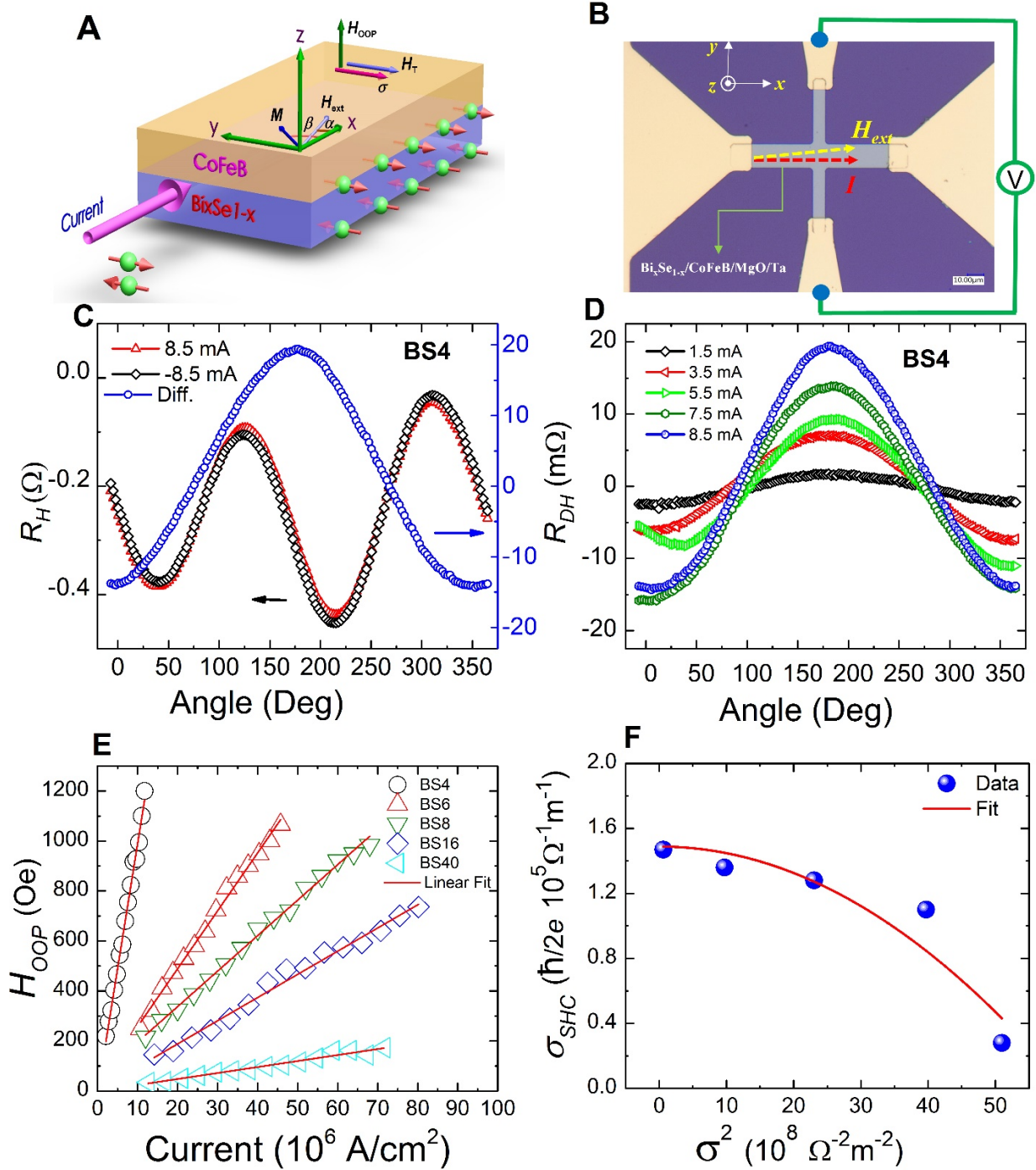

**Fig. 3. Current-induced magnetization switching in the Bi$_x$Se$_{(1-x)}$ (4 nm)/Ta (0.5 nm)/CoFeB(0.6 nm)/Gd (1.2 nm)/CoFeB (1.1 nm) heterostructure.** **(A)** A schematic drawing of the switching sample stack structure. **(B)** The $R_{AHE}$ measured in the Bi$_x$Se$_{(1-x)}$ switching sample using a 50 μA current. **(C)** and **(D)** Current-induced switching of the magnetization due to the SOT arising from the Bi$_x$Se$_{(1-x)}$ underlayer in the presence of a constant 8 mT in-plane bias field measured in Hall cross bar with dimensions 15 μm × 70 μm.

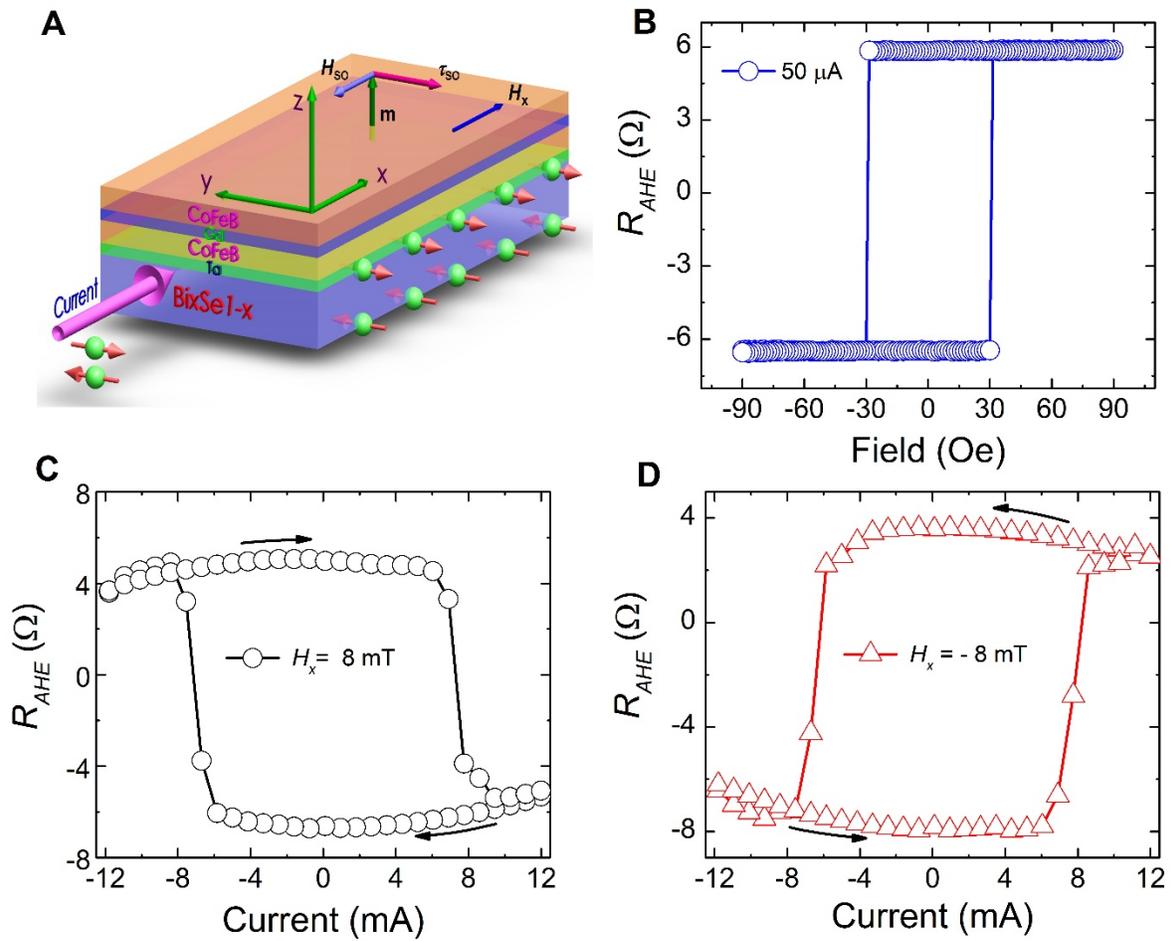

**Table 1.** A summary of the $\theta_{SH}$, $\sigma_{SH}$, $J_{sw}$, and $\sigma$ of the SHMs in this work and best previous reports.

| Parameters | Bi$_x$Se$_{(1-x)}$ (This work) (RT) | (Bi$_{0.5}$Sb$_{0.5}$)$_2$Te$_3$ (Ref. 13) (1.9 K) | Bi$_2$Se$_3$ (Ref. 12) (RT) | $\beta$-Ta (Ref. 6) (RT) | $\beta$-W (Ref. 17) (RT) | Pt (Ref. 7, 9) (RT) |
|---|---|---|---|---|---|---|
| $\sigma$ ($\Omega^{-1}$m$^{-1}$) | $0.78 \times 10^4$ | $2.2 \times 10^4$ | $5.7 \times 10^4$ | $5.3 \times 10^5$ | $4.7 \times 10^5$ | $4.2 \times 10^6$ |
| $\sigma_{SH}$ ($10^5 \frac{\hbar}{2e} \Omega^{-1}$m$^{-1}$) | 1.5 | 93.5 | 2.0 | 0.8 | 1.9 | 3.4 |
| $\theta_{SH}$ | 18.83 | 425 | 3.5 | 0.15 | 0.4 | 0.08 |
| $J_{sw}$ (A/cm$^2$) | $2.3 \times 10^5$ | $8.9 \times 10^4$ | -- | $5.5 \times 10^6$ | $1.6 \times 10^6$ | $2.85 \times 10^7$-$10^8$ |